\documentclass[10pt,a4paper]{article}

\usepackage[a4paper,top=2cm,bottom=2cm,left=2cm,right=2cm]{geometry}

\usepackage{amsmath,amssymb,amsfonts,graphicx,pifont,dcolumn}
\usepackage{color}
\usepackage{cancel}
\usepackage{ulem}
\usepackage{comment}
\RequirePackage{slashed}
\usepackage{cite}
\usepackage{hyperref}

\title{Scalar spectrum in a graviton soft wall model}

\author{Matteo Rinaldi$^1$ \\ \\ $^1$Universit\`a degli studi di 
Perugia, 
Dipartimento di Fisica 
e Geologia.  \\ INFN Sezione di Perugia. Via A. Pascoli, 06123, 
Perugia,  Italia.
\\~
\\
 Vicente Vento$^2$ \\ \\ $^2$ 
Departamento de F\'{\i}sica Te\'orica-IFIC, Universidad de
Valencia- CSIC, \\ 46100 Burjassot (Valencia), Spain.}

\date{}

\begin{document}

\maketitle

\begin{abstract}

In this study 
we present a unified phenomenological analysis of the scalar 
glueball and  scalar meson spectra within an AdS/QCD 
framework in the bottom up approach. 
For this purpose we generalize the recently developed graviton 
soft-wall (GSW) model, which has shown
an excellent agreement with the lattice QCD glueball spectrum, to a  
description of glueballs and mesons with a unique energy scale. In this scheme, 
dilatonic effects, are 
incorporated in the metric as a deformation of the AdS space.
We apply the model also to the heavy meson spectra with success. We obtain
quadratic mass equations for all scalar mesons while the glueballs satisfy 
an almost linear mass equation.
Besides their spectra, we also discuss  the
 mixing of scalar glueball and light scalar meson states 
 within a unified framework: the GSW model. To this aim, the 
light-front holographic  approach, which connects the
 mode functions of AdS/QCD  to the light-front wave functions,  is 
applied.
 This relation provides the probabilistic interpretation required
 to properly investigate the mixing conditions.

\end{abstract}



\section{Introduction}

In the last few years,
hadronic models, inspired by  the holographic conjecture 
\cite{Maldacena:1997re,Witten:1998zw}, have been vastly used and developed in order 
 to investigate 
non-perturbative features of glueballs and mesons, thus trying to 
grasp fundamental features  of QCD 
\cite{Fritzsch:1973pi,Fritzsch:1975wn}. Recently we have 
used the so 
called 
AdS/QCD models to study the glueball 
spectrum~\cite{Vento:2017ice,Rinaldi:2017wdn}. The holographic principle relies in a correspondence between a 
five dimensional 
classical theory with an AdS metric and a supersymmetric
 conformal quantum 
field theory with $N_C  \rightarrow \infty$. This theory, different from QCD, is taken as a starting point  to construct a 5 dimensional 
holographic dual of it. This is the so called bottom-up approach~\cite{Polchinski:2000uf,Brodsky:2003px,DaRold:2005mxj,Karch:2006pv}. 
To do so,  models are constructed by modifying  the five dimensional  
classical AdS theory with the aim of resembling  QCD as much as 
possible.
The main differences characterizing these models are related to the 
strategy used to break conformal invariance.  

{ It must be noted that the 
predictions  for observables in these models are at leading 
order in the number of colours expansion.}
For example, the meson masses are 
$\mathcal{O}(N^0_c)$, thus these models reproduce the essential features of 
the meson spectrum~
\cite{Erlich:2005qh,Colangelo:2008us,deTeramond:2005su}.
For mesons and baryons, the AdS/QCD 
approaches have been also successfully used to describe form factors and 
various types of parton distribution 
functions~\cite{deTeramond:2005su,Rinaldi:2017roc,Bacchetta:2017vzh,
deTeramond:2018ecg}. 
Within this formalism, also the glueball masses, being 
$\mathcal{O}(N_C^0)$, have been 
studied~\cite{Rinaldi:2017wdn,Colangelo:2007pt,Capossoli:2015ywa}. 
Besides these developments, which are in line with the 
present investigation, other models have been introduced recently  by
using the
bottom-up holography. For example, an interesting development is the 
no-wall model 
\cite{Afonin:2013npa}, which has been successful in explaining the 
heavy 
quark spectra.

{ In this investigation, we start from the holographic
Soft-Wall (SW) model, were a dilaton field  
is introduced to softly break conformal invariance, which allows to properly reproduce 
the Regge trajectories  of the meson spectra.
Within this scheme we have recently  introduced a model, the graviton soft-wall model (GSW)~\cite{Rinaldi:2017wdn,Rinaldi:2018yhf}, which
has been able  to describe the lattice QCD glueball spectrum~\cite{Morningstar:1999rf,Chen:2005mg,Lucini:2004my}, 
whose slope was not reproduced by the traditional SW models. 
Since one of the goals of the present analysis is a fully 
characterization of the mixing between glueball and scalar meson 
spectra, it 
is fundamental to consider hadronic models which provide a complete 
description of 
both spectra.  Therefore, in here we generalize the GSW model to study also the scalar mesons, known as the  
$f_0$ mesons~\cite{Patrignani:2016xqp,Tanabashi:2018oca},
particles with spin parity $J^{PC}= 0^{++}$. Our approach provides 
in fact an excellent description of the light and heavy meson spectra 
and is able to reproduce properly 
their Regge trajectories.}

 In the next section we discuss the bottom-up approach 
of the AdS/QCD correspondence~\cite{Polchinski:2000uf,Brodsky:2003px,DaRold:2005mxj,Karch:2006pv},
and describe the  generalization of the GSW model \cite{Rinaldi:2017wdn} to describe both the glueballs and the mesons 
in the same model and with a unique energy scale. In section III 
we discuss glueball-meson mixing and finally in the conclusions
we extract some consequences of our analysis.

\section{Scalar glueball and light scalar meson spectrum in the graviton soft-wall model}

The GSW model describes quite well the scalar 
glueball spectrum~\cite{Rinaldi:2017wdn} of quantum gluodynamics (QGD). 
However, the conventional SW  models based on the $AdS_5$ metric do not lead to a good simultaneous description of the 
glueball and meson spectra.  We show next  that if we generalize  the 
metric by incorporating an exponential factor $e^{\alpha \varphi(z)}$ 
the 
GSW model achieves that goal.
 These kind of modifications have been adopted in several 
improvements of the SW model, see Refs. 
\cite{Colangelo:2007pt,Capossoli:2015ywa,last,Gutsche:2019blp,
Vega:2016gip } . 
The GSW model is defined by the metric,

\begin{equation}
ds^2=\frac{R^2}{z^2} e^{\alpha \varphi(z)} (dz^2 + \eta_{\mu \nu} 
dx^\mu dx^\nu) = \frac{R^2}{z^2} e^{\alpha \varphi(z)} g_{M N} dx^M 
dx^N  =\bar{g}_{MN}dx^M dx^N.
\label{metric5}
\end{equation}
The function $\varphi(z)$ will be specified later on and the need for 
$\alpha$ will become apparent in the next subsections.
 This type of metrics with different prefactor $\alpha$ have been used 
by many authors up to very recently to explain the properties and 
spectra of hadrons ~\cite{ 
Klebanov:2004ya,MartinContreras:2019kah,FolcoCapossoli:2019imm,
Bernardini:2016qit,Li:2013oda,Vega:2016gip}.
Quantities evaluated in the GSW model are displayed with overline. 
The relation between the standard $AdS_5$ metric and 
$\bar{g}_{MN}$ is

\begin{eqnarray}
\bar g^{MN} &= &e^{-\alpha \varphi(z)} g^{MN}, \\
\sqrt{-\bar{g} } & = & e^{\frac{5}{2}\alpha \varphi(z)} \sqrt{-g}.
\end{eqnarray}

The GSW action in the scalar sector is  then defined by

\begin{equation}
\bar{\mathit{I}} = \int d^4x dz \sqrt{-\bar{g}} e^{\beta \varphi(z)} 
\big[\bar{g}^{MN} \partial_M S(x,z) \partial_N S(x,z) + M^2_{5 m} 
S^2(x,z) \big],
\label{GSWaction}
\end{equation}
where $R^2 M^2_{5 m} = -3$ is the $AdS_5$ mass of the scalar meson, 
$S(x,z)$ the scalar meson field and $e^{\beta \varphi(z)}$ is a dilaton 
used to describe the soft-wall behaviour.
 Here $x$ stands for Minkowski coordinates and $z$ for the coordinates in the fifth dimension.
 In terms of the standard 
$AdS_5$ metric, this action becomes 

\begin{equation}
\label{action}
\bar{\mathit{I}} = \int d^4x dz\sqrt{-g} e^{\varphi(z)(\frac{3}{2}\alpha 
+\beta)}  \big[g^{MN} \partial_M S(x,z) \partial_N S(x,z) + 
e^{\alpha \varphi(z)} M^2_{5 m} S^2(x,z)\big].
\end{equation}
 
Given these equations, we obtain the equation of motion (EoM) for the 
scalar { field}. The EoM of the scalar 
mesons and its spectrum is determined by variation of the action Eq. 
(\ref{action})  with respect to the scalar field. 
{ The profile function of the dilaton field is the same of that 
adopted in the standard SW model:}

\begin{equation}
 \varphi(z) =  k z^2.
 \end{equation}
where $k$ is a scale factor that is determined by fitting 
the spectra. { Such a functional form leads to a correct description 
of
the Regge trajectories of the meson spectra.}
 The parameter
  $\beta$ is fixed by imposing that
$\frac{3}{2}\alpha + \beta = -1$.
Such a relation ensures that
 the kinetic term for the 
scalar meson  is the same of that described in the standard
 SW model ~\cite{Colangelo:2008us,Capossoli:2015ywa}.
This ansatz was crucial 
to reproduce the Regge behaviour in the meson 
sector~\cite{Rinaldi:2017wdn,Karch:2006pv,Erlich:2005qh}. 
 
In the present analysis, we start by fitting the lattice glueball spectrum 
with the mass equation obtained from the Einstein equation for 
the graviton~\cite{Rinaldi:2017wdn}. The fit will fix the product $\alpha k$. We next proceed
to fit the PDG scalar meson spectrum with the mass equation derived from
the  EoM of the scalar 
field. This fit leads to separate values for $\alpha$ and $k$.

\subsection{Glueballs}

 The Einstein equation,  for the metric Eq. (\ref{metric5}), leads 
to the glueball mode equation in the
 5th-variable $z$ once the $x$ dependence has been factorized  as 
$\Phi(z) e^{i x_\mu q^\mu}$, where $q^2 = -M^2$ and $M$ represents the 
mass of the glueball modes
 
 \begin{equation}
\label{7}
 \frac{d^2 \Phi(z)}{dz^2} - \left(\alpha k z + \frac{3}{z}\right)
 \frac{d \Phi 
(z)}{d z} +  \left(\frac{8}{z^2} - 6 \alpha k - 4 \alpha^2 k^2 z^2 + M^2\right) 
\Phi (z) - \frac{8}{z^2} e^{\alpha k z^2} \Phi (z) = 0~,
 \end{equation}
By performing the change of function

\begin{equation}
\Phi (z) = e^{\alpha k z^2/4} 
\left(\frac{z}{\alpha k}\right)^{\frac{3}{2}} \phi (z)
\end{equation}
we get  a Schr\"odinger type equation

\begin{equation}
- \frac{d^2 \phi(z)}{d z^2} + \left(  \frac{8}{z^2} e^{\alpha k z^2} -\frac{15}{4} \alpha^2 k^2 z^2 +
7 \alpha k - \frac{17}{4 z^2} 
\right) \phi(z) = M^2 \phi(z).
\end{equation}
In this equation it is apparent that $M^2$ represent the mode mass squared which will arise from the eigenvalues of an 
{\it Hamiltonian} operator scheme. 

It is convenient to move to the adimensional variable $ 
t=\sqrt{\alpha k/2}\; z$ and we define the mode by 
$\Lambda^2 = (2 /\alpha k)\; M^2$. The the equation becomes

\begin{equation}
-\frac{d^2 \phi(t)}{d t^2} + \left(\frac{8}{t^2} e^{2 t^2} - 15 t^2 + 14 
- \frac{17}{4 t^2}  \right) \phi(t) = \Lambda^2 \phi(t).
\label{Gexact}
\end{equation}
This is a typical Schr\"odinger equation with no free parameters except 
for an energy scale in the mass determined by $\alpha k$. The 
potential term is uniquely determined by the metric and only the scale 
factor is unknown and will be determined from  lattice QCD.
This equation has no exact solutions and numerical solutions can be 
found \cite{Rinaldi:2017wdn}. One can be tempted to make the 
approximation

\begin{equation}
\frac{e^{2 t^2}}{t^2} \sim \frac{1}{t^2} + 2 + 2 t^2,
\end{equation}
which leads to
\begin{equation}
-\frac{d^2 \phi(t)}{d t^2} + \left( t^2 + 30 + \frac{15}{4 t^2}  \right) 
\phi(t) = \Lambda^2 \phi(t),
\label{Gapprox}
\end{equation}
which is a Kummer equation that  has exact solutions whose spectrum is given by
$$\Lambda^2_n = 4 n + 36, \; n= 0, 1, 2,\ldots.$$ 
{ The above approximation could in principle be relevant since it 
leads to a Schr\"odinger equation similar to that obtained  
 in the SW model, see e.g. Ref. 
\cite{Capossoli:2015ywa}. In addition, in the same Ref. 
\cite{Capossoli:2015ywa} 
such an approximation has been directly used to solve the EoM  for the 
scalar glueball, dual to the scalar field, if a modified AdS space is 
considered. However, at variance with Ref. \cite{Capossoli:2015ywa}, 
in our approach we considered the graviton as the dual to the glueball 
field \cite{Rinaldi:2017wdn}. Therefore, the EoM, Eq. (\ref{Gexact}),  
is 
different from that 
described and solved in Ref. \cite{Capossoli:2015ywa} and does not allow 
an approximate analytical solution. In fact, as 
reported   }
in Table \ref{Gmodes}, we see that the result from the exact and 
approximate calculation are quite different, 

\begin{table} [htb]
\begin{center}
\begin{tabular} {|l c c c c c|}
\hline
$n$ & 0 & 1 & 2 & 3 & $\ldots$ \\
\hline
$\Lambda^2_n$ (exact) & 53.88 & 82.17 & 117.02 & 157.95& $\ldots$ \\
\hline
$\Lambda^2_n$ (approx)& 36 & 40 & 44 & 48 & $\ldots$ \\
\hline
\end{tabular}  
\caption{Exact versus approximate glueball modes.}
\label{Gmodes}
\end{center}
\end{table}

\begin{figure}[htb]
\begin{center}
\includegraphics[scale= 0.85]{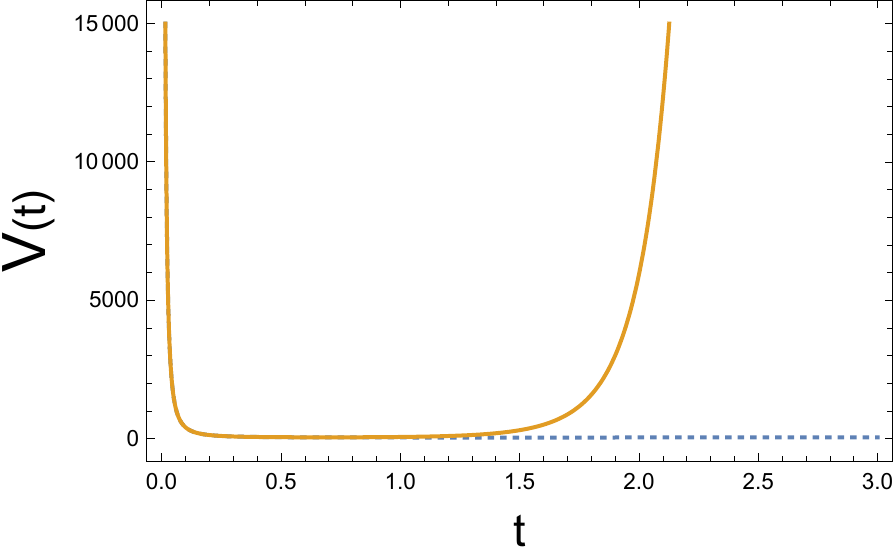} \hskip 0.5cm 
\includegraphics[scale= 0.84]{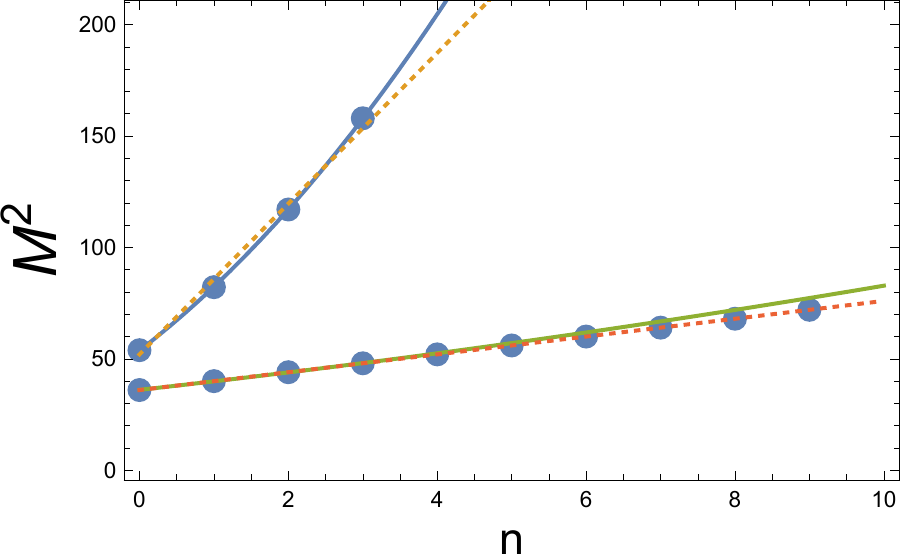}
\end{center}
\caption{Left: We plot as a function of $t$ the potentials of the 
equivalent exact (solid), Eq.(\ref{Gexact}), and approximate (dotted), 
Eq.(\ref{Gapprox}), Schr\"odinger equations. Right: We plot the 
glueball 
modes of the exact solution and approximate solution noticing that the 
  latter is an almost linear relation with  $M^2$ 
while the former is 
quadratic in mass in the region of our interest.}
\label{Gpotentials}
\end{figure}

The numerical solution is very different from the approximate one 
as  it is clear from the potentials shown in the left panel of 
Fig. 
\ref{Gpotentials}. One can see that for the approximate solution, the 
potential rises only for very large values of $t$ and therefore the 
bound states are close to each other in energy, while the potential for 
exact solution rises for $t \sim 2 $ thus the bound states are 
well separated in energy. Moreover, we see in  the right panel of 
Fig. \ref{Gpotentials} that the behaviour of the modes is also very 
different. Finally we want to stress that
in the exact solution the dependence of the mass  is linear with $n$ 
in 
the low $n$ region while in the approximation solution the mass squared 
is linear in $n$.

\subsection{Mesons}

In the case of the scalar mesons the variation of the action, Eq. 
(\ref{action}), leads to

\begin{equation}
\partial_M(\sqrt{-g} e^{-\varphi(z)} g^{MN} \partial_N S(x,z)) =  
\sqrt{-g}  e^{-\varphi(z)(1-\alpha)}  M^2_{5 m}  S(x,z).
\end{equation}
Once we separate the $x$ dependence by factorizing $S(x,z) = \Sigma(z) 
e^{-i q_\mu x^\mu}$ with $q^2 = -M^2$, where  $M$ is the mass of the 
meson modes, we get

\begin{equation}
- \frac{d^2 \Sigma(z)}{dz^2} +\left(\frac{3}{z} + 2 kz\right) \frac{d 
\Sigma(z)}{dz} + \frac{3}{z^2} e^{\alpha \varphi(z)} \Sigma(z) = M^2 
\Sigma(z)
\end{equation}
where we have substituted the value of the scalar $AdS_5$ mass. By 
setting $\varphi (z)= k z^2 $ and 
performing the change of function 

\begin{equation}
\Sigma(z) =\left( \frac{z}{k}\right)^\frac{3}{2} e^{ k z^2/2} 
\sigma(z),
\end{equation}
we get a Schr\"odinger type equation

\begin{equation}
-\frac{d^2 \sigma(z)}{dz^2} + \left( k^2z^2 + 2 k + \frac{15}{4z^2} 
- \frac{3}{z^2} e^{\alpha k z^2} \right) \sigma(z) = M^2 \sigma(z).
\end{equation}

We now proceed to the change to the adimensional variable $ u= 
\sqrt{k/2}\; z$,

\begin{equation}
-\frac{d^2 \sigma(u)}{du^2} + \left( 4 u^2 + 4 + \frac{15}{4u^2} - 
\frac{3}{u^2} e^{2 \alpha u^2} \right) \sigma(u) =  \Omega^2 \sigma(u),
\label{MAexact}
\end{equation}
where $\Omega^2= (2/k)M^2 $.

Let us perform the same approximation as before, namely to expand the 
exponential and keep up to three terms to obtain 

\begin{equation}
-\frac{d^2 \sigma(u)}{du^2} + \left(  (4 - 6 \alpha) u^2 + (4 - 6 
\alpha) + \frac{3}{4u^2} \right) \sigma(u) =  \Omega^2 \sigma(u).
\label{Mapprox}
\end{equation}

This equation can be transformed into a Kummer type equation by the 
change of variables $v= (4- 6 \alpha^2)^{1/4}u$ 

\begin{equation}
-\frac{d^2 \sigma(v)}{d v^2} + \left(v^2 + \frac{4 - 6 \alpha}{\sqrt{4 - 
6 \alpha^2}} + \frac{3}{4 v^2}\right) \sigma(v) = 
\frac{\Omega^2}{\sqrt{4 - 6 \alpha^2}},
\label{Mkummer}
\end{equation}
which has an exact spectrum given by 

\begin{equation}
\Omega_n^2 = 4(n+1) \sqrt{4 - 6 \alpha^2} + 4 - 6\alpha, \; n = 
0,1,2,\ldots.
\label{Mmodes}
\end{equation}
and the mode functions are

\begin{equation}
\sigma(v) = \mathcal{N} e^{-v^2/2} v^{3/2} { _1F_1}(- n, 2, v^2)
\label{kummerf}
\end{equation}
where $\mathcal{N}$ is a normalization factor and $_1F_1$ is a well 
known hypergeometric function and recall that $v = (4-6\alpha^2)^{(1/4)} 
\; u $
where $u = (\sqrt{k/2} \;z)$. Note that the approximate solution only 
has bound states for $|\alpha| < \sqrt{2/3}$. 

\begin{figure}[htb]
\begin{center}
\includegraphics[scale= 0.85]{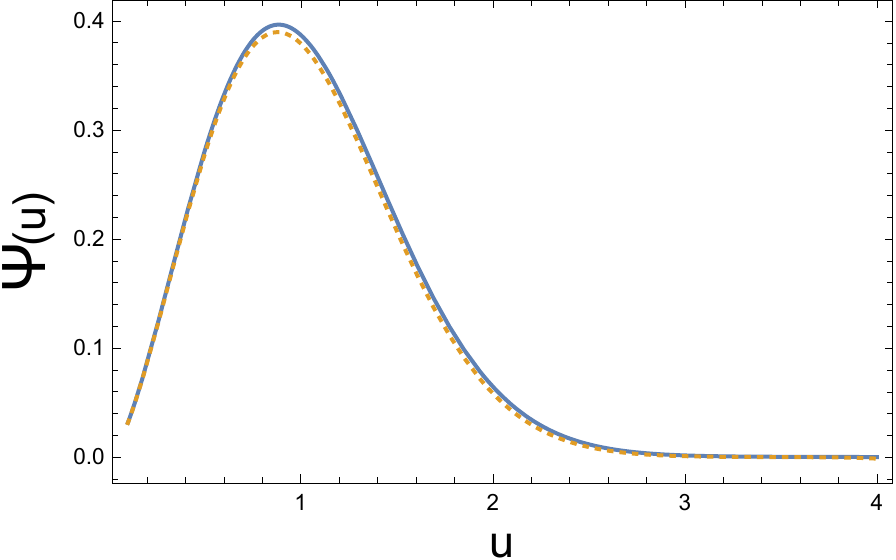} \hspace{0.5cm} 
\includegraphics[scale= 0.87]{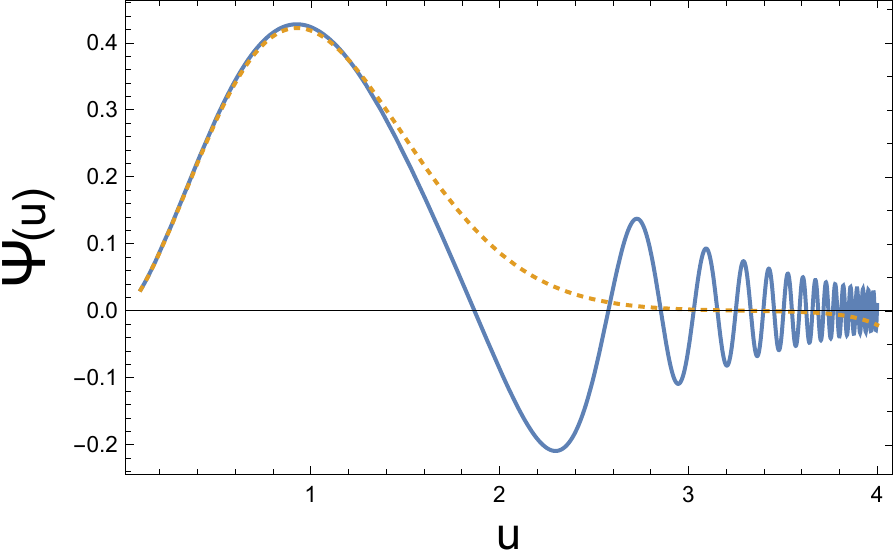}
\end{center}
\caption{We plot as a function of $u$  for $\alpha =0.2$ (left) and 
$\alpha=0.4$ (right)  the exact solution for the lowest mode (solid), 
Eq.(\ref{MAexact}), and the approximate lowest mode solution (dotted) 
obtained by expanding in the exact mode equation the exponential up to 
third order.}
\label{Msolutions}
\end{figure}

The meson modes  are a function of $\alpha$. In the next section we 
will 
proceed to fix the parameters of the model by using phenomenological 
inputs. In the left panel of Fig. \ref{Msolutions} we show the exact 
lowest mode 
for $\alpha=0.2$ (solid)
together with the approximate solution obtained by expanding the 
exponential up to the third order (dotted).  For small values of 
$\alpha$ both solutions are very similar and their mode values { 
are} almost equal. For larger values of $\alpha =0.4$ (see right panel 
of Fig. \ref{Msolutions}), closer to the no binding limit, the mode 
values are very similar but the exact solution becomes unstable. We 
shall discuss these details further in the next section.

\subsection{Phenomenological analysis}

Our AdS/QCD model provides us with a succession of mass modes of 
differential equations
 which should be numerically solved. 
The equation for the glueballs Eq.(\ref{Gexact}) depends on a scale 
$\alpha k$ 
which we will use to match the glueball spectrum obtained from lattice 
QCD in the quenched approximation which is Gluodynamics 
~\cite{Morningstar:1999rf,Chen:2005mg,Lucini:2004my}. The equation for 
the mesons, Eq.(\ref{MAexact}), depends on $k$ and $\alpha$, { 
separately}. We  use the 
PDG spectrum to fit the mesons 
spectrum~\cite{Patrignani:2016xqp,Tanabashi:2018oca}, which is 
certainly 
QCD, if we assume that QCD is the theory of the strong interactions. In 
principle, the scale of the glueballs and the mesons should be the same 
if we could  fit QCD data, but since we are using lattice QCD in the 
quenched approximation to fit the latter, one could expect some minor 
differences between the two. We will choose the same mass scale. The 
fitting  procedure is different from that used before by many 
authors~\cite{Rinaldi:2017wdn,Colangelo:2008us,Capossoli:2015ywa} were 
an overall scale was used since in this case the scale affects the mode 
functions, which correspond to the light cone wave 
functions\cite{Brodsky:2003px}, through the relation between the $t$ 
and 
$z$ variables.

\begin{figure}[htb]
\begin{center}
\includegraphics[scale= 1.0]{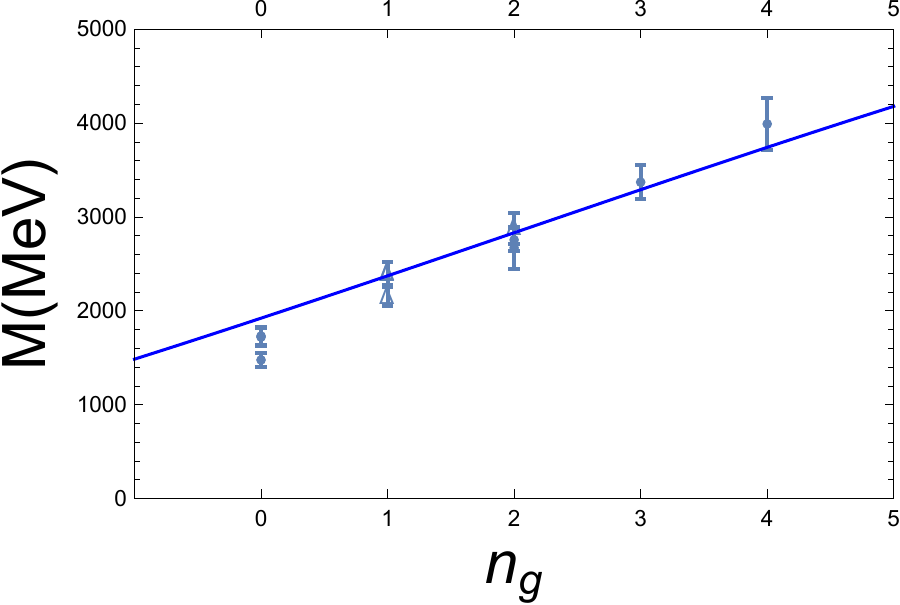} 
\end{center}
\caption{We plot the average lattice values for the scalar and 
tensor glueball masses of refs. 
\cite{Morningstar:1999rf,Chen:2005mg,Lucini:2004my} as a function of 
mode number. We show the fit to the spectrum from the modes of Eq. 
\ref{Gexact} by using just one parameter $\alpha k$.} 
\label{GlueballFit}
\end{figure}

We proceed to fit the glueball spectrum. The used  lattice data are 
shown in Table~\ref{Gmasses} 
~\cite{Morningstar:1999rf,Chen:2005mg,Lucini:2004my}   
\footnote{We have not included the lattice results from the unquenched 
calculation 
~\cite{Gregory:2012hu} to be consistent, which however,  in this
range of
masses and for these quantum numbers are in agreement, within errors, 
with the
shown results.}.
We also  use for the fit  the results for the tensor 
glueball states since the theory predicts degeneracy between the scalar 
and the tensor glueball for all soft-wall models. The value of the 
parameter, obtained in the fit shown in Fig. \ref{GlueballFit}
with the GSW model, is $\alpha k =  (370$ MeV)$^2$.
One should notice that the fit to the scalar glueball spectrum obtained 
by solving  exactly the EoM leads to an almost linear
relation between the mass and the mode number ~\cite{Rinaldi:2017wdn}.

\begin{table} [htb]
\begin{center}
\begin{tabular} {|c c c c c c c|}
\hline
$J^{PC}$& $0^{++}$&$2^{++}$&$0^{++}$&$2^{++}$&$0^{++}$&$0^{++}$\\
\hline
MP & $1730 \pm 94$ & $2400 \pm122$ & $2670 \pm 222 $&  & &  \\
\hline
YC & $1719 \pm 94$ & $2390 \pm124$ &  &  &  &  \\
\hline
LTW & $1475 \pm 72$ & $2150 \pm 104$ & $2755 \pm 124$& $2880 \pm 
164 $& $3370
\pm 180$& $3990 \pm 277$  \\
\hline
\end{tabular}  
\caption{Glueball masses [MeV] from lattice calculations by MP
~\cite{Morningstar:1999rf}, YC~\cite{Chen:2005mg} and LTW 
~\cite{Lucini:2004my}.}
\label{Gmasses}
\end{center}
\end{table}

\begin{table} [htb]
\begin{center}
\begin{tabular} {|c c c c c c c c c|}
\hline
Meson& $f_0(500)$ 
&$f_0(980)$&$f_0(1370)$&$f_0(1500)$&$f_0(1710)$&$f_0(2020)$&$f_0(2100)$&
$f_0(2200)$\\
\hline
PDG & $475 \pm 75$ & $990 \pm 20$ & $1350 \pm 150 $&$1504 \pm 6 $  & 
$1723 \pm 6 $&  $1992 \pm 16 $&  $2101 \pm 7 $&$ 2189 \pm 13$\\
\hline
\end{tabular}  
\caption{Scalar meson masses [MeV]  from 
PDG~\cite{Patrignani:2016xqp,Tanabashi:2018oca}.}
\label{Mmasses}
\end{center}
\end{table}

Now that $\alpha k$ has been fixed, we will use $\alpha$ to fit the 
meson spectrum. In order to estimate the value of $\alpha$
let us use the approximate equation Eq.(\ref{Mkummer}).
From the mode equation Eq.(\ref{Mmodes}) we get the spectrum

\begin{equation}
 M_n^2 = \Omega_n^2 \dfrac{k}{2}=\Omega_n^2 \dfrac{(370)^2}{2 
\alpha}~ = A(\alpha) n + B(\alpha),
\label{mesonmass}
\end{equation}
where $A(\alpha) = (740)^2\, (\sqrt{4-6 \alpha^2}/2 \alpha)$ and $ 
B(\alpha) = (740)^2 \;((\sqrt{4-6 \alpha^2} + 1 - \frac{3}{2} \alpha)/2 
\alpha)$.
Thus in the meson case our complicated equation of motion leads to a 
quadratic relation between mass and mode number, which is not the case 
for glueballs.

\begin{figure}[htb]
\begin{center}
\includegraphics[scale= 0.63]{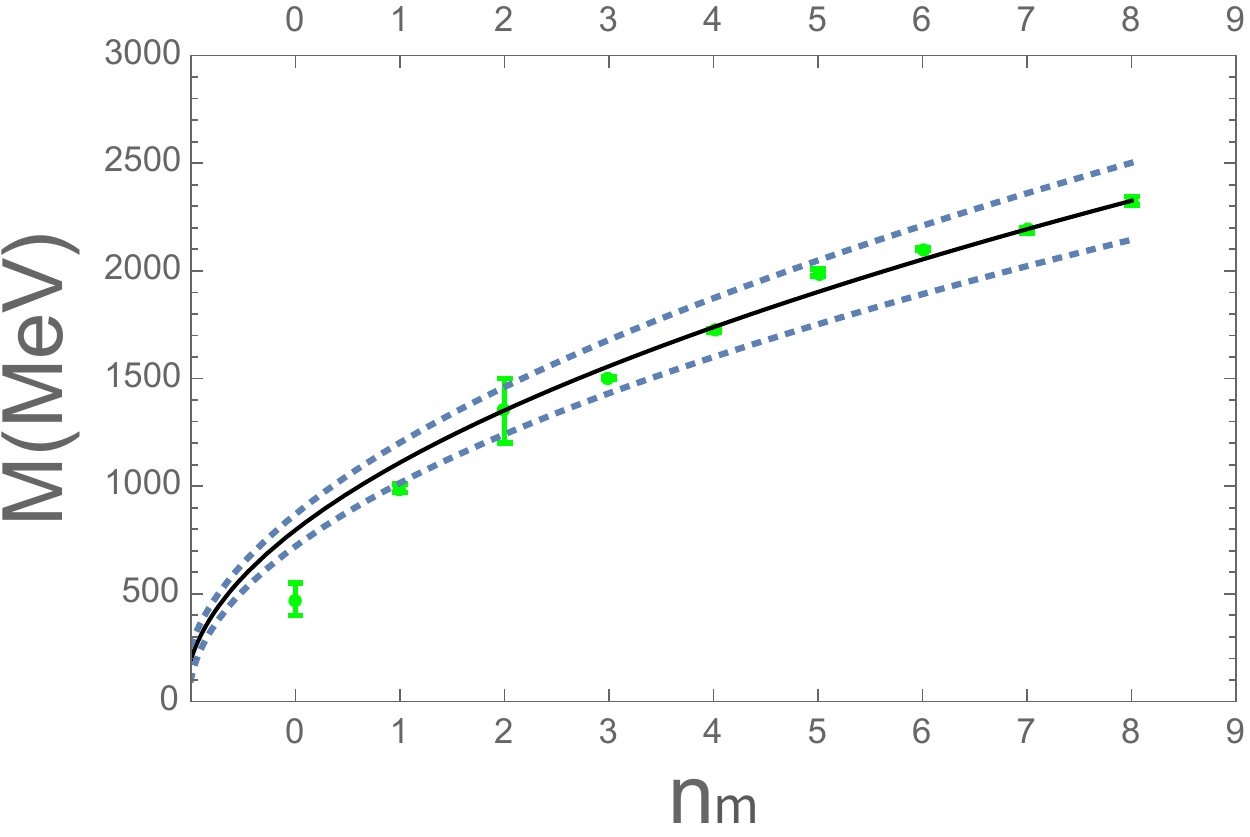} 
\hspace{0.5cm}\includegraphics[scale= 0.63]{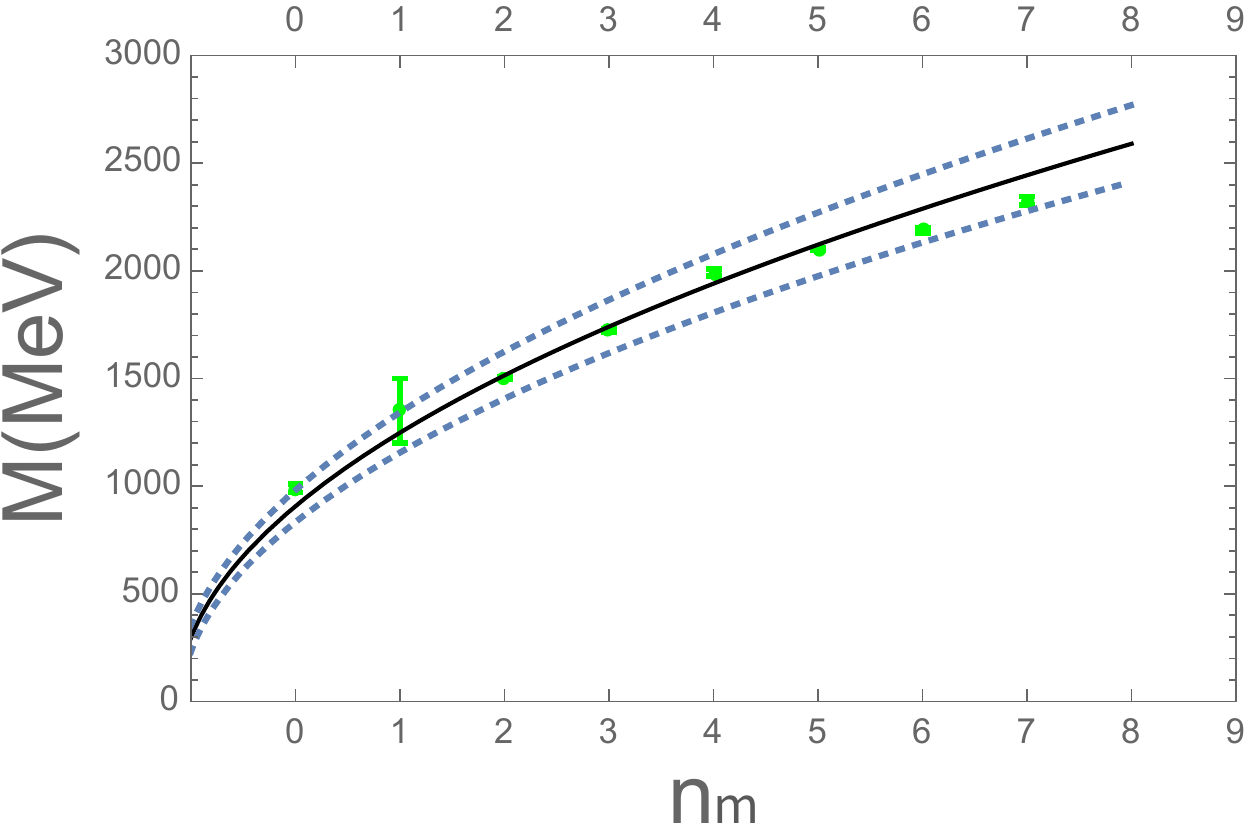} 
\end{center}
\caption{We plot the $f_0$ PDG meson 
spectrum~\cite{Patrignani:2016xqp,Tanabashi:2018oca}
as a function of mode number. Left panel: including the $f_0(500)$. The 
fits 
correspond to $\alpha=0.61$ (solid) and $\alpha=0.61 \pm 0.04$ 
(dotted). Right panel: 
without the $f_0(500)$. The fits correspond to $\alpha=0.55$ (solid) 
and $\alpha=0.55\pm0.04$ (dotted).} 
\label{MesonFit}
\end{figure}

Having this analytic equation the fitting procedure is quite 
straightforward. The first question is if we should include the 
$f_0(500)$ in the fit. As discussed in our 
work~\cite{Rinaldi:2018yhf}, many authors have argued { that this 
} is not 
a conventional meson state but a tetraquark or a 
hybrid~\cite{Mathieu:2008me,Tanabashi:2018oca}. In the present GSW 
model we have less freedom since
the energy scale is fixed by the glueballs. In Fig.~\ref{MesonFit}
 we
 fit the PDG meson spectrum shown in Table~\ref{Mmasses} with 
our model. 
The left figure includes  the $f_0(500)$ while the right figure does 
not.

It is apparent again that the $f_0(500)$ is difficult to fit if the 
mass is so low as given in the PDG tables, and therefore, probably, the 
dynamical mechanism forming it is different from the rest of the 
mesons. 
From now on we will not consider the $f_0(500)$ in our fits to the 
meson 
spectrum.

\begin{figure}[htb]
\begin{center}
\includegraphics[scale= 1.0]{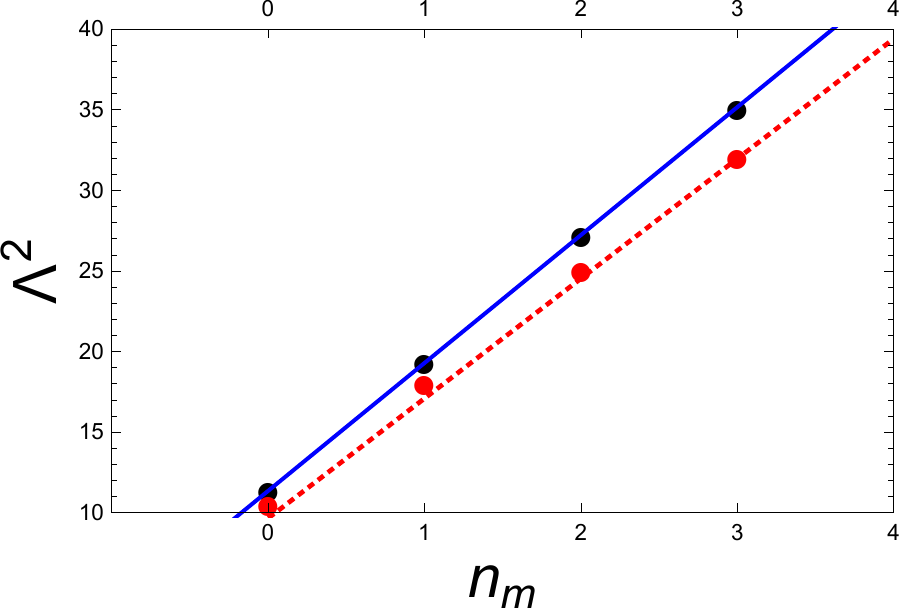} 
\end{center}
\caption{The dots represent the mode values of the true solution. 
The upper points correspond to $\alpha=0.1$, the lower points to 
$\alpha=0.2$. The lines represent the mode values of the approximate 
solution. The solid curve for $\alpha=0.1$ the dotted curve for 
$\alpha=0.3$.} 
\label{Extrapolation}
\end{figure}

The problem with the true solution, that corresponding to 
Eq.(\ref{MAexact}), is that it is very unstable as the binding 
potential becomes weaker, i.e. for values of $\alpha > 0.2$ and much 
more unstable for the higher modes as shown in the right panel of Fig. 
\ref{Msolutions}. 
However, the spectrum does no differ much from the approximate spectrum 
and the mode functions have a similar structure before the large $u$ 
oscillations appear. 
One should notice that the value for which the approximate solution 
starts to be different from the exact solution, $u>1.5$, corresponds to 
$ z> 10/\Lambda_{QCD}$. Thus for a strong confined system such region 
is 
of little relevance.
In Fig. \ref{Extrapolation} we analyse the behaviour 
of the mode values of the true solution. The dots represent the true 
solution. The upper points have been calculated for $\alpha=0.1$  and 
the lower points for $\alpha = 0.2$. The lines represent the mode 
values 
of the approximate solution for $\alpha=0.1$ (solid) and $\alpha = 0.3$ 
(dotted). Thus we see an almost perfect fit for $\alpha=0.1$ but for 
$\alpha=0.2$ the higher modes of the exact solution tends towards 
higher 
values of $\alpha$ of the approximate solution. If we look back to the 
right plot of  Fig. \ref{MesonFit} this is exactly what is happening, 
the lower values of the meson spectrum are fitted quite well by the 
approximate solution for $\alpha= 0.51$, while the upper modes are 
fitted by the solution for $\alpha=0.59$. Having in mind these caveats 
from now on we will work with the approximate equation plotting not 
single curves but bands which take into account the difference between 
the true solution and the approximate one. The latter can be 
considered as the theoretical error in our approach.

One may conclude from the above analysis that the GSW model
 describes well both the scalar glueball and 
meson spectra with the same scale $\alpha k$. Let us remark 
that the energy scale arises from the metric as does the
$\alpha$ modification which builds up the mesons. It is 
worth noticing that the value of $\alpha$ obtained from our 
phenomenological analysis for mesons is very close to the value of 
$\alpha$ obtained in Ref. \cite{FolcoCapossoli:2019imm} from a 
theoretical analysis developed for glueballs.
{ In addition the dilatonic constant $\sqrt{k} = 
0.5$ GeV is extremely close to the values reported in Refs. 
\cite{deTeramond:2018ecg,Forshaw:2012im}, i.e. $M_\rho/\sqrt{2}$.}
 Let us recall that this fit
 is  $\mathcal{O}(N_C^0)$ and that corrections 
$\mathcal{O}(N_C^{-1})$ and higher should be added to obtain a 
precise value. However, 
the fact that the fit is quite good suggests that the contribution of
 the higher terms might be small.
 
  Now that we have shown that our model describes well both 
spectra,
a comment is necessary to highlight the importance of the structure of 
the metric. 
We have seen that the glueball spectrum in the region under scrutiny 
here has an almost
 linear relation between the mass and the mode number and this
 result is produced
 by our equations of motion (\ref{7}) 
which cannot be approximated by a 
 Kummer type equation as previously discussed.
 On the other hand, the meson equation, which is quite similar in 
structure to that of the glueball equation,
 acquires an $\alpha$ in the exponent of the exponential potential term. 
The value of $\alpha$ is
 crucial to allow for binding, but in so doing transforms the mode 
equation into an almost
 Kummer type equation with a quadratic relation between the mass and the 
mode number, as
 the experimental data indicate.

Now that we have an equation for the scalar mesons in the 
light sector, is it possible to generalize 
the equation to incorporate the heavy masses. In order to reproduce the 
heavy spectrum we have to add 
to the dynamics the mass of the heavy quarks. Several 
 procedures are available
\cite{Branz:2010ub,Kim:2007rt,Afonin:2013npa} and we choose at this 
moment a very simple ansatz, namely to add a constants to the mass 
equation, which will be proven to be successful,

\begin{equation}
M_n = \sqrt{A(\alpha) n + B(\alpha)} + C,
\label{heavymass}
\end{equation}
where $C$ is the contribution of the quark masses, thus there will be a 
$C_c$ for the $c \bar{c}$ states and a different one $C_b$ for
the $ b \bar{b}$ states. In table \ref{heavymasstable} we show the 
values of the heavy scalar mesons masses collected by PDG 
~\cite{Patrignani:2016xqp,Tanabashi:2018oca}. It must be noted that 
some 
of the states do not have confirmed quantum numbers and others 
 states themselves need to be confirmed. 

\begin{table} [htb]
\begin{center}
\begin{tabular} {|c c c c c c c c c|}
\hline
&& &Scalar& Heavy &Mesons &&&\\
\hline $c \bar{c} $ &$ 
\chi_{c0}(1P)$&$\chi_{c0}
(3860)$&$X(3915)$&$X(3940)$&$X(4160)$&$X(4350)$& 
$\chi_{c0}(4500)$&$\chi_{c0}(4700)$
\\
\hline
$I^G(J^{PC})$& 
$0^+(0^{++})$&$0^+(0^{++})$&$0^+(0/2^{++})$&$?^?(?^{??})$&$?^?(?^{??}
)$&$0^+(?^{??})$&$0^+(0^{++})$ &$0^+(0^{++})$\\
\hline
 PDG & $3414 \pm 0.30$ & $3862^{+66}_{-45}$ & $3918 \pm 1.9 $&   
$3942^{+13}_{-12}  $&  $4156^{+40}_{-35} $&  $4350^{+5.3}_{-5.1} $ & 
$4506^{+42}_{-41}  $ & $4704^{+24}_{-34} $\\
\hline
 $b \bar{b}$& $ \chi_{b0}(1P)$&$ \chi_{b0}(1P)$& & & & & & \\
\hline
$I^G(J^{PC})$& $0^+(0^{++})$&$0^+(0^{++})$& & & & & &  \\
\hline
 PDG & $9859 \pm 0.73$ & $9912.21 \pm 0.57$ &  &  &  & & & \\
\hline
\end{tabular}  
\caption{Heavy scalar meson spectrum [MeV] from PDG 
\cite{Tanabashi:2018oca}. Notice that some  of the particles are only 
suspected to be scalars and others need confirmation.}
\label{heavymasstable}
\end{center}
\end{table}

\begin{figure}[htb]
\begin{center}
\includegraphics[scale= 0.85]{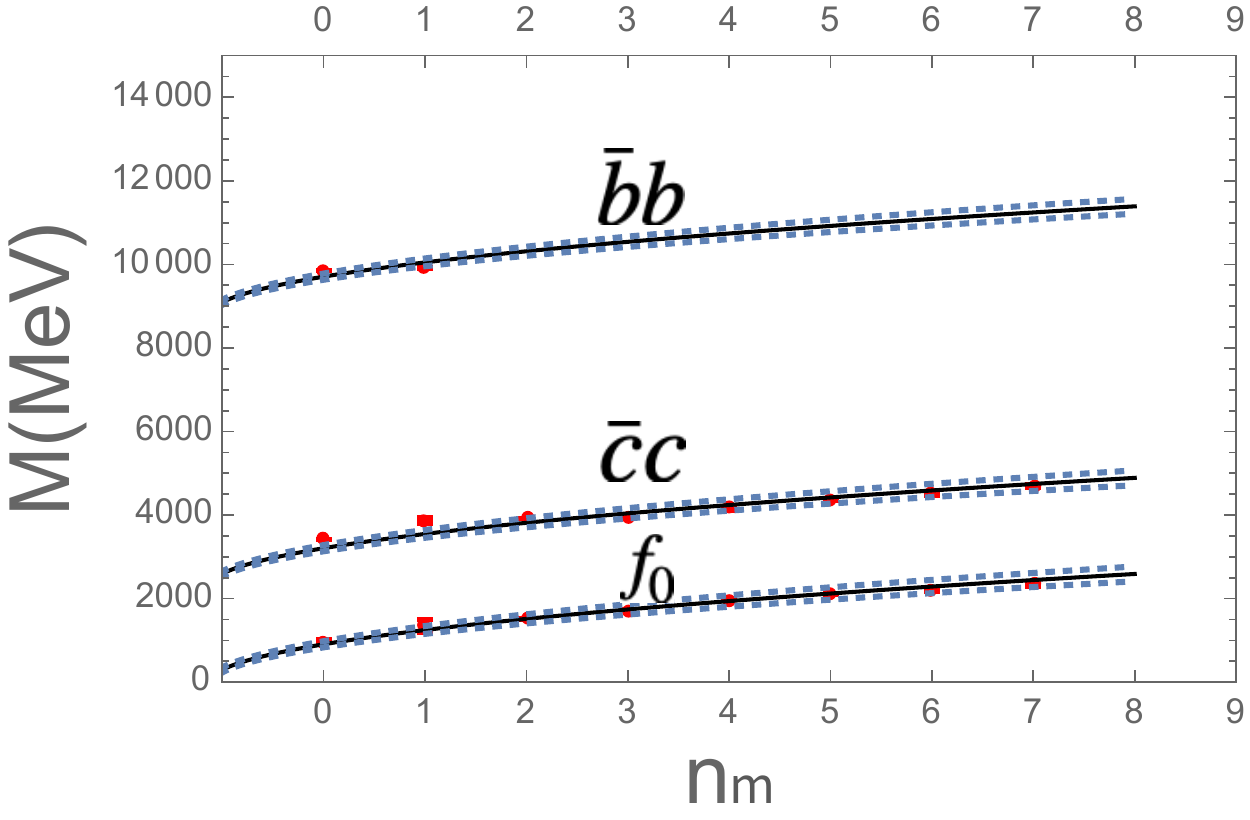}  
\end{center}
\caption{We show the scalar meson spectrum for all quark sectors. The 
light dots 
represent the scalar meson spectrum obtained from experimental data 
~\cite{Patrignani:2016xqp,Tanabashi:2018oca}. The  curves correspond to 
Eq. (\ref{heavymass}) with $C_c = 2400$ MeV for the $c \bar{c}$ mesons 
and $C_b = 8700$ MeV for the $b \bar{b}$ mesons and the same two values 
of $\alpha$, i.e.  $0.55$ (solid) and $0.55\pm0.04$ (dotted) for all 
mesons.}
\label{heavyspectrum}
\end{figure}

We show in Fig. \ref{heavyspectrum} the fit obtained by 
Eq.(\ref{heavymass}) with the following parametrization. 
We have kept $\alpha$ fixed to the same values for all mesons, i.e. 
$\alpha=0.55$ (solid) and $\alpha=0.55\pm0.04$ (dotted) and we have 
chosen for 
the light quark systems $C=0$,   
$C_c=2400$ MeV, for the $ c \bar{c}$ mesons, and for the $ b \bar{b} $ 
mesons $C_b=8700$ MeV. The fits are 
excellent given the simplicity of the model and moreover, and this is 
very exciting, $ C_c \sim 2 m_c$ and  $C_b \sim 2 m_b$. The heavy 
mesons 
have not been used at all to fit $\alpha$ and the model is showing that 
within its simplicity it is capable to reproduce, quite nicely, the 
scalar meson 
spectrum and therefore giving a lot of credibility to the glueball 
spectrum { and the GSW model itself}. Moreover we see that all 
mesons satisfy approximately the same 
mass trajectories apart from an overall scale associated with the quark 
masses and that all the elements in the PDG suspected of being scalar 
mesons seem to be scalar mesons. The model has proven to be 
tremendously 
predictive.

\section{Glueball-Meson mixing}

\begin{figure}[htb]
\begin{center}
\includegraphics[scale= 0.82]{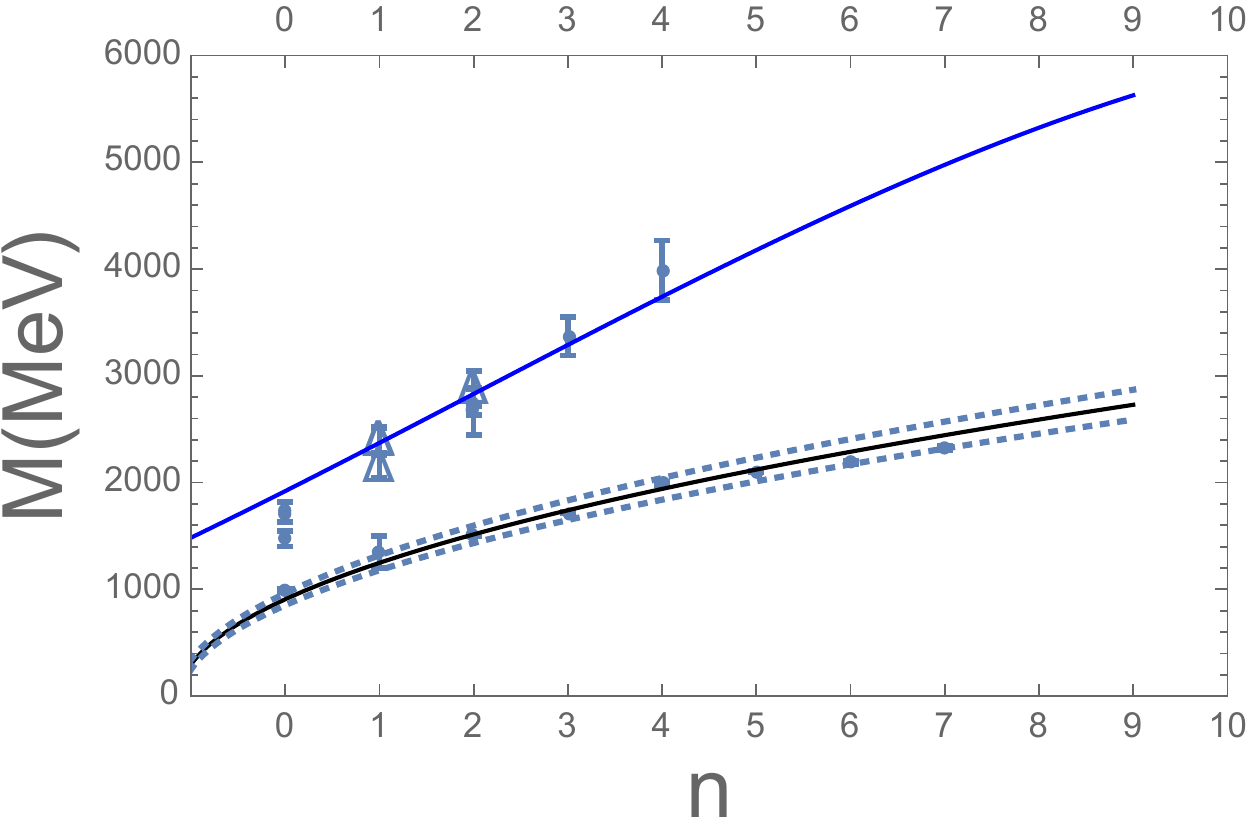}  
\end{center}
\caption{We show fits to the glueball spectrum (solid upper line) and 
scalar 
meson spectrum (lower band) obtained within the GSW model. The dark dots 
represent 
glueball  spectrum obtained from lattice 
QCD~\cite{Morningstar:1999rf,Chen:2005mg,Lucini:2004my}. The light dots 
represent the scalar meson spectrum obtained from experimental data 
~\cite{Patrignani:2016xqp,Tanabashi:2018oca}. We have extended the fits 
to high mode numbers to motivate our analysis of mixing.}
\label{SpectrumFit}
\end{figure}

{ In this section we provide an interpretation of the Mixing 
mechanism within a unified holographic description. In fact, as 
described in several analyses, see, e.g., Refs. 
\cite{Mathieu:2008me,Vento:2015yja,Amsler:1995td,Close:2002zu,
Chanowitz:1982qj}, glueballs states can mix with mesons with similar 
masses and quantum numbers. Such an effect prevents a clear and safe 
extraction of the glueball spectra from data. It is therefore 
fundamental to provide a detailed description of the mixing effect in 
order to identify kinematic conditions where glueball states 
disentangle 
from those of mesons.}
 In Fig.~\ref{SpectrumFit} we show  the glueball lattice data 
(upper points) and the meson data (lower points). We plot also the fits 
with the GSW model discussed before and extend the fits to higher mode 
numbers to find  that  glueball masses with a certain mode number are 
equal to  meson masses with a much larger mode number. For example, the 
glueball masses for $n_g=0,1,2$ are similar to the scalar meson masses 
for $n_m=4,7,10$ respectively. The difference in mode numbers grows as 
the masses of the glueballs increase due to the different slopes of the 
fitting curves. This observation has led us  to discuss 
the meson glueball mixing scenario for high hadron 
masses in a recent paper~\cite{Rinaldi:2018yhf}. We proceed to analyze 
the consequences of 
that observation in { unified framework, by using} the GSW model.

Scalar glueballs might mix with scalar 
mesons~\cite{Mathieu:2008me,Vento:2015yja}.
Recently,  in view of the spectra of mesons and glueballs in AdS/QCD 
models, we have discussed the possibility 
that at high energies mixing might not be favorable and states with 
mostly gluonic valence structure might exist \cite{Rinaldi:2018yhf}. 
This is an exciting possibility since the  presence of almost pure 
glueball states and the study of their decays would help in 
understanding many properties of QCD related to the physics of gluons. 
In here we have described an AdS/QCD model in the scalar sector that 
describes both scalar mesons an scalar glueballs on an equal footing by 
means of a unique energy scale. This energy scale enters the 
description 
of the mode functions which can be interpreted, in the previously 
developed scheme, as a light cone wave functions. 

 In order to investigate mixing between meson and glueball 
states we  make use, in this section, of the Light-Front (LF) AdS/QCD 
formalism, see Ref. \cite{Zou:2018eam} for a useful review. In this 
formalism  the equations of motion which describe the propagation of  
modes in anti-de Sitter space is taken to be equivalent to the equation 
for the light-front wave function for hadrons~\cite{Brodsky:2006uqa}. 
This approach has been successfully applied to describe several 
observable related to the non perturbative  features of QCD such as 
decay constants, form factors, parton distribution functions (PDFs), 
generalised PDFs, transverse momentum dependent PDF and gravitational 
form factors 
\cite{Brodsky:2006uqa,Brodsky:2007hb,Brodsky:2008pf,deTeramond:2008ht,
Brodsky:2010ur,Vega:2009zb,Forshaw:2012im,Chakrabarti:2017teq,
Rinaldi:2017roc,Bacchetta:2017vzh}. As shown in \cite{Brodsky:2010ur}, 
the LF Holography matches the QCD running coupling in both the 
perturbative and non perturbative regions. Moreover, also the 
description of the $\rho$ meson electroproduction via LF AdS/QCD  
matches the data~\cite{Forshaw:2012im}. Due to all these remarkable 
results, we proceed to use LF AdS/QCD to describe the LF wave functions 
of glueballs and mesons calculated in terms  of  the 
solutions of 
the mode equations in the gravity sector

The 
holographic light-front 
representation of the equation of
motion, in $AdS$ space can be recast in the form 
of a light-front 
Hamiltonian~\cite{Brodsky:2003px}

\begin{equation}
H_{LC} |\Psi_n> = M^2 |\Psi_n>.
\end{equation}
In the AdS/QCD light-front framework the above relation becomes 
a Schr\"odinger type equation

\begin{equation}
 \left(-\frac{d^2}{d t^2} + V(t) \right) \Psi(t) = \Lambda^2 \Psi(t)
 \label{Sch}
 \end{equation}
 where $t$ and $\Lambda^2$ in this equation are adimensional.
 The holographic light-front wave function are defined by $\Psi_n(t) 
 =\; <t|\Psi_n>$ and are normalized
as

\begin{equation}
<\Psi_n|\Psi_n> = \int dt |\Psi_n(t)|^2 = 1
\label{prob}
\end{equation}
The described mode functions introduce, in  this way,  
probability distributions.

The eigenmodes of Eq.(\ref{Sch}) determine the mass spectrum.
In this framework, we get functions of $t$ for which we can define a 
probability distributions as in Eq.(\ref{prob}).
In ref. \cite{Rinaldi:2018yhf} we discussed the mixing in a two 
dimensional Hilbert space generated by  a meson 
and a glueball states, \{$|\Psi_m>, |\Phi_g>$\}.
   We found out 
that the mixing probability is proportional to the overlap probability 
of these two wave functions, i.e  $|<\Psi_m|\Phi_g>|^2$ .
Thus, in order to discuss the mixing, we need the mode functions. In 
the case 
of the glueballs we  find the mode functions numerically since we have 
shown that the approximate solution is very different from the exact 
one, while in the meson case we  use the approximate solution, which 
behaves similarly, with some caveats, to the exact mode function. 
The meson equation can be brought into  Kummer's form 
Eq.(\ref{Mkummer}) 
by a convenient change of variables in terms of which the the mode 
functions are given by Eq.(\ref{kummerf}).

\begin{figure}[htb]
\begin{center}
\includegraphics[scale= 0.8]{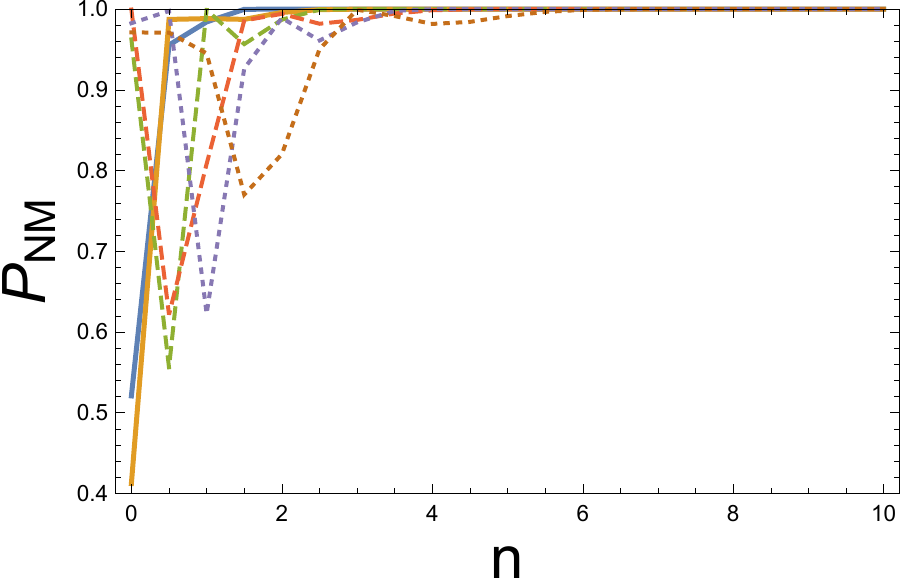} \hskip 0.3cm
 \includegraphics[scale= 0.8]{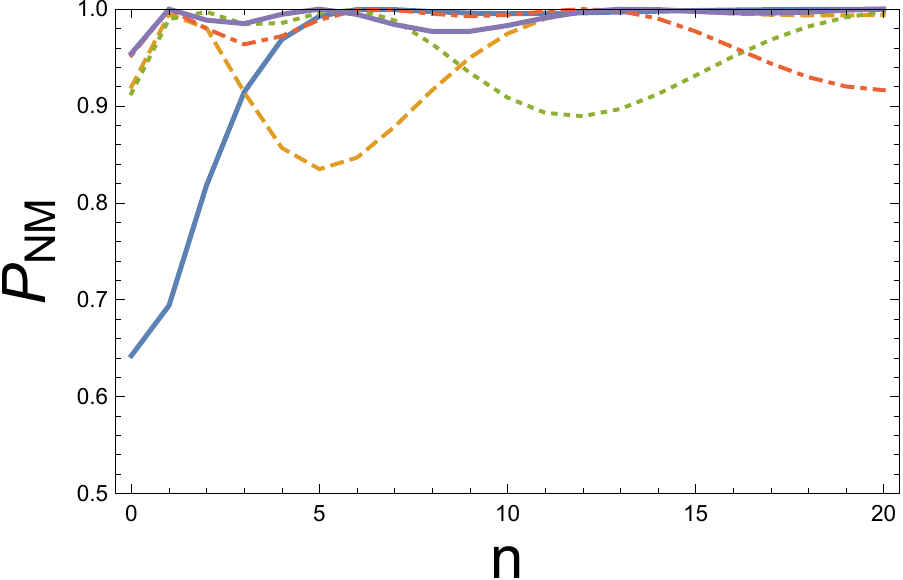}
\end{center}
\caption{We plot the probability of no mixing 
as function of the meson mode number, $n$, and fixed glueball mode 
numbers. Left panel: $n_g = 0$ (solid), $1$ (dashed), $3$ 
(dotted).
{  Results have been obtained when both
glueball and meson wave functions have been calculated within the 
same  GSW model. We plot two lines for each $n_g$, one calculated with 
the meson wave function for $\alpha=0.51$ and the other calculated with 
the 
wave function  for $\alpha=0.59$. Right 
Panel: $n_g=0$ (solid), $1$ (dashed), $2$ (dotted), $3$
(dot-dashed), $4$ (solid). Results have been obtained by using the GSW 
model to evaluate the  
glueball w.f. and the SW model for that of the  meson, see Ref. 
\cite{Rinaldi:2018yhf}.}}
\label{Overlap1}
\end{figure}

\begin{figure}[htb]
\begin{center}
\includegraphics[scale= 1.1]{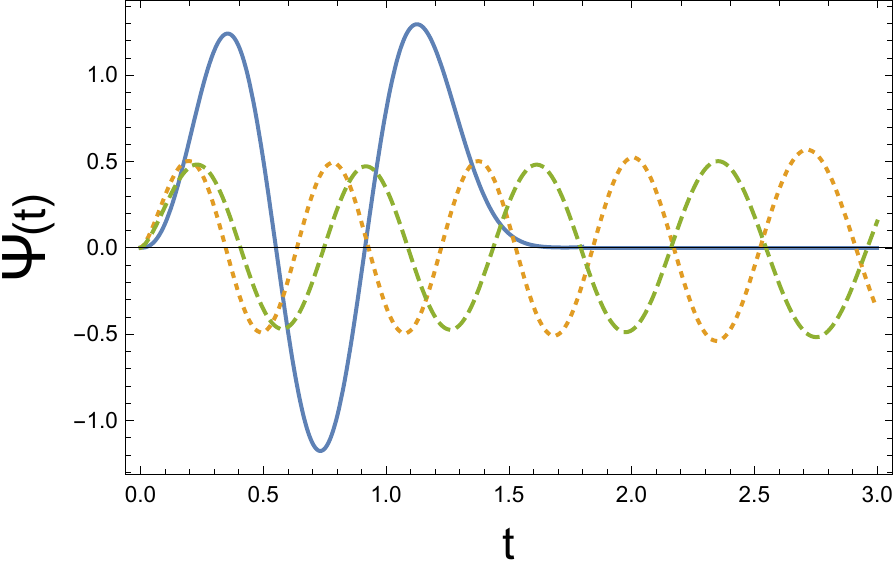} 
\end{center}
\caption{ We plot the glueball mode function for $n_g =2$ and the meson 
mode functions for $n_m=10$ and $\alpha=0.51$ (dashed) and $\alpha = 
0.59$ (dotted). }
\label{Overlapwaves}
\end{figure}

In Fig. \ref{Overlap1} we  plot the probability of no mixing for low lying glueball modes ($n_g=1,2,3$) overlapping with mesons of  modes up to mode number $n_m = 10$. We observe that the overlap probability is small for the larger meson mode numbers and that it is only sizeable when the mode numbers of the two states are not very different. Let us make the discussion more detailed with an example. We choose a glueball of mode number $n_g=2$
and a meson of mode number $n=10$. They  can be 
considered as candidates for  mixing because  the glueball  mass in the model is $m(n_g =2) \sim 2800$ 
MeV  and the meson mass is also $m(n_m =10) \sim 2800 $ 
MeV. Thus this example is a prototype for a mixing scenario 
for heavy particles.   In Fig.\ref{Overlapwaves} we show the
 mode functions 
for the $n_g=2$ glueball mode and that for the $n_m =10$ meson mode for 
the GSW model as a function of $t$.   We notice that the meson function 
oscillates rapidly due to its larger mode number and therefore the 
overlap integral becomes very small. Note that the true mode function 
will be in between those two drawn, closer to the one of shorter 
wavelength for low modes and to the one of longer wavelength for the 
higher modes.

Looking at Figs.~\ref{SpectrumFit} and left panel of Fig. 
\ref{Overlap1}, we see that
a favorable mixing scenario is mostly excluded in the case of heavy 
glueballs and mesons, since the  mass condition is satisfied for very 
different mode numbers. For example it can be seen that for
 $n_g=2,3,4$ the favorable meson modes 
of almost equal masses occur for  $n \sim 10,13,17$.
{ By comparing the left and right panels of Fig. \ref{Overlap1}, one 
 can  notice that only a unified description of glueballs and 
mesons within the same holographic model predicts a highly picked probability of mixing 
for low mode numbers and an extremely small probability for mode numbers 
that differ in several units. In fact, once the unified GSW model is 
adopted,  for $n_g \leq 3$ then the  mixing probability is negligible 
for $n_m \geq 3$.  On the contrary, if one uses two different models, 
a tail at large $n_m$ can be found.    }
The outcome of our analysis is that the AdS/QCD approach 
predicts the existence of almost pure 
glueball states in the scalar sector in the mass range above $2$ GeV.

\section{Conclusion}
In this work we have discussed the application of Graviton Soft Wall 
Model for both 
 scalar glueballs and  mesons. To this aim, 
we have initially compared the theoretical spectrum   with lattice $QCD$ 
data
 in the case of the 
glueballs and the experimental $f_0$ spectrum of the PDG tables in 
the case of the mesons.  The model 
introduces a unique energy scale for both glueballs and mesons.
We have shown that the model respects the Regge behaviour
and nicely reproduces both these spectra at leading order in $1/N_c$.
How Regge behaviour comes about for the mesons is 
critically dependent on the value of the metric
as described by the parameter $\alpha$.
 In our approach, the intrinsic Regge behaviour of the meson 
spectrum is a direct consequence of the metric and on the specific value of 
the parameter $\alpha$.
 The non linear equation obtained 
for the spectrum changes from a linear dependence on the mass to a 
quadratic dependence on the mass with the value of this parameter. In 
the latter case the non linear equation can be approximated by a Kummer 
type equation while in the former not. Thus the beauty of the model 
resides in that with similar equations it is able to reproduce the 
almost linear slope of the glueball spectra and the quadratic Regge 
trajectory for mesons. Moreover, the value of $\alpha$, obtained by 
fitting the meson spectrum $ \sim 0.55 \pm 0.04$ is very similar to 
that 
 obtained by theoretical arguments from the glueball spectrum 
\cite{FolcoCapossoli:2019imm}.
In addition, we remark that also the dilatonic constant we found is 
extremely close to that reported in Refs. 
\cite{deTeramond:2018ecg,Forshaw:2012im}.
We have extended the model to the  heavy quarks sector simply by 
adding a constant related to the quark masses to our equation for the 
spectrum as justified by other authors. The result of the analysis is 
excellent. Thus, the model is able to explain, within the limits of the 
leading $1/N_C$ dominance, all the scalar sector and moreover, predicts 
that some particles, whose quantum numbers are not known or  need 
confirmation, are scalars or tensors.

We have noted that in our model the slope of the glueball 
spectrum, as 
a function of mode 
number, is also bigger than that of the meson spectrum 
and therefore for heavy almost degenerate glueballs and light quark mesons 
states their mode numbers differ considerably.  Assuming a light-front 
quantum mechanical description of AdS/QCD correspondence, we have shown 
that the overlap probability of these heavy glueballs to heavy light quark mesons
 is small and 
thus one expects little mixing in the relatively high mass sector. Therefore, 
this is the kinematical region to look for almost pure glueball states.
At present stage, large statistics of Central Exclusive Process (CEP) 
data
is being collected by the LHC experiments, and we expect exciting new 
results to appear concerning the higher mass gluon enriched process.

\section*{Acknowledgments}
We acknowledge Sergio Scopetta and Marco Traini for
discussions. VV thanks the hospitality extended to him by the 
University of Perugia  and the INFN group in Perugia. This work was supported in part 
by MICINN and UE Feder under contract FPA2016-77177-C2-1-P, and by the 
STRONG-2020 project of the European Unions Horizon 2020 research and 
innovation programme under grant agreement No 824093.

\end{document}